\documentclass[manuscript]{aastex631}
\usepackage{CJK}
\usepackage{graphicx}
\usepackage{epstopdf}
\usepackage{multirow}
\usepackage{ulem}

\shorttitle{harmonic emissions driven by the loss-cone distribution}
\shortauthors{Ning et al.}

\graphicspath{{./}{figures/}}

\newcommand\wpe{\omega_\mathrm{pe}}
\newcommand\wce{\Omega_\mathrm{ce}}
\newcommand\Rone{\uppercase\expandafter{\romannumeral1}}
\newcommand\Rtwo{\uppercase\expandafter{\romannumeral2}}
\newcommand\Rthree{\uppercase\expandafter{\romannumeral3}}

\begin{document}
\begin{CJK*}{UTF8}{gbsn}
\title{Harmonic maser emissions from electrons with the loss-cone distribution in solar active regions}

\correspondingauthor{Yao Chen}
\email{yaochen@sdu.edu.cn}

\author{Hao Ning (宁昊)}
\affiliation{Institute of Space Sciences, Shandong University, Shandong, 264209, People's Republic of China}
\affiliation{Institute of Frontier and Interdisciplinary Science, Shandong University, Qingdao, Shandong, 266237, People's Republic of China}

\author{Yao Chen (陈耀)}
\affiliation{Institute of Frontier and Interdisciplinary Science, Shandong University, Qingdao, Shandong, 266237, People's Republic of China}
\affiliation{Institute of Space Sciences, Shandong University, Shandong, 264209, People's Republic of China}

\author{Sulan Ni (倪素兰)}
\affiliation{Institute of Space Sciences, Shandong University, Shandong, 264209, People's Republic of China}
\affiliation{Institute of Frontier and Interdisciplinary Science, Shandong University, Qingdao, Shandong, 266237, People's Republic of China}

\author{Chuanyang Li (李传洋)}
\affiliation{Institute of Frontier and Interdisciplinary Science, Shandong University, Qingdao, Shandong, 266237, People's Republic of China}
\affiliation{Institute of Space Sciences, Shandong University, Shandong, 264209, People's Republic of China}

\author{Zilong Zhang (张子龙)}
\affiliation{Institute of Space Sciences, Shandong University, Shandong, 264209, People's Republic of China}
\affiliation{Institute of Frontier and Interdisciplinary Science, Shandong University, Qingdao, Shandong, 266237, People's Republic of China}

\author{Xiangliang Kong (孔祥良)}
\affiliation{Institute of Space Sciences, Shandong University, Shandong, 264209, People's Republic of China}
\affiliation{Institute of Frontier and Interdisciplinary Science, Shandong University, Qingdao, Shandong, 266237, People's Republic of China}

\author{Mehdi Yousefzadeh}
\affiliation{Institute of Space Sciences, Shandong University, Shandong, 264209, People's Republic of China}
\affiliation{Institute of Frontier and Interdisciplinary Science, Shandong University, Qingdao, Shandong, 266237, People's Republic of China}
%
%




\begin{abstract}


Electron cyclotron maser emission (ECME) is regarded as a plausible source for the coherent radio radiations from solar active regions (e.g., solar radio spikes). In this Letter, we present a 2D3V fully kinetic electromagnetic particle-in-cell (PIC) simulation to investigate the wave excitations and subsequent nonlinear processes induced by the energetic electrons in the loss-cone distribution. The ratio of the plasma frequency to the electron gyrofrequency $\wpe/\wce$ is set to 0.25, adequate for solar active region conditions. As a main result, we obtain strong emissions at the second-harmonic X mode (X2). While the fundamental X mode (X1) and the Z mode are amplified directly via the electron cyclotron maser instability, the X2 emissions can be produced by the nonlinear coalescence between two Z modes and between Z and X1 modes. This represents a novel generation mechanism for the harmonic emissions in plasmas with a low value of $\wpe/\wce$, which may resolve the escaping difficulty of explaining solar radio emissions with the ECME mechanism.

\end{abstract}


\section{Introduction} \label{sec:intro}

Solar radio bursts with brightness temperatures much higher than the effective temperature of the emitting electrons are caused by coherent radiation of plasmas. Plasma emission (PE) and electron cyclotron maser emission (ECME) were proposed as two major mechanisms for the coherent emissions, occurring in plasmas with the frequency ratios (ratio of plasma frequency to electron gyrofrequency, $\wpe/\wce$) larger and less than unity, respectively \citep[see,][and references therein]{Melrose...2017}. In solar active regions with strong magnetic fields, plasmas can be characterized by $\wpe/\wce<1$, and ECME is the favored mechanism \citep[see, e.g.,][]{Regnier...2015}.

Energetic electrons with positive gradients of velocity distribution function (VDF) can excite plasma waves via the electron cyclotron maser instability (ECMI). In plasmas with $\wpe/\wce <1$, ECMI can amplify escaping emissions in extraordinary or ordinary mode (X/O mode) directly, known as ECME. The framework of the ECME theory was first proposed by \cite{Twiss...1958} and \cite{Schneider...1959}. \cite{Wu&Lee...1979} greatly advanced the theory by taking both the relativistic correction and the doppler-shift term into account. Follow-up studies have applied the theory to radio emissions of various kinds of astrophysical sources \citep[see reviews by][]{Wu...1985,Treumann...2006}. Among various types of solar radio bursts, millisecond spikes, characterized by high brightness temperatures, short durations, and narrow bandwidths \citep[see, e.g.,][]{Droege...1977,Benz...1985,Benz...1986,Huang&Nakajima...2005,Feng.et.al.2018,Karlicky.et.al.2021}, have been attributed to ECME.

One major problem of ECME when applied to solar radio bursts is about how fundamental emissions ($\omega \approx \wce$) pass through the second-harmonic layer in the solar corona without being absorbed considerably \citep{Melrose&Dulk...1982}. Ideas suggested to circumvent the problem include: propagation through a density-depleted tunnel \citep[e.g.,][]{Wu.et.al.2014,Melrose&Wheatland...2016}, escape through transmission windows at $\theta =0^\circ$ or $90^\circ$ \citep[where $\theta$ is the angle between the directions of wavevector and background magnetic field]{Robinson...1989}, and re-emission after being absorbed by the electrons \citep{McKean.et.al.1989}.

Another promising possibility is to generate harmonic emissions ($\omega  \approx n\wce$, $n\ge 2$) that can escape without passing through the second-harmonic absorption layer. The latest studies employing fully kinetic electromagnetic particle-in-cell (PIC) simulations have found that harmonic emissions in X2 can be amplified by energetic electrons with VDFs manifesting strip-like features or horseshoe-like anisotropy \citep{Ning.et.al.2021, Yousefzadeh.et.al.2021}.

For solar magnetic loops, trapped energetic electrons can develop a loss-cone type VDF, which has been considered as the most favored driver of ECME. The loss-cone ECME for solar spikes has been studied over decades \citep[see, e.g.,][]{Holman.et.al.1980, Melrose&Dulk...1982, Winglee&Dulk...1986,Aschwanden...1990, Stupp...2000, Lee.et.al.2013}. According to existing literature, in plasmas with $\wpe/\wce<1$ the dominant modes generated by the loss-cone ECME are fundamental X and O modes (X1 and O1) and the non-escaping Z mode, while the growth of harmonic emissions is insignificant.

It has not been yet resolved whether electrons with a loss-cone type VDF can generate efficient harmonic emissions in plasmas with $\wpe/\wce<1$. Early studies mainly calculate the growth rates of relevant modes based on the linear kinetic theory. Some researchers developed quasi-linear theory to study the growth and saturation of amplified waves by considering the quasi-linear diffusion of the VDF \citep[e.g.,][]{Aschwanden...1990,Yoon&Ziebell...1995,Ziebell&Yoon...1995}. PIC simulations have been performed since the 1980s in one-dimensional \citep[e.g.,][]{Wagner.et.al.1983,Wagner.et.al.1984} or two-dimensional domains \citep{Pritchett.et.al.1999}. However, most studies were conducted under relatively small numbers of macroparticles per cell and mainly focused on the linear stage of wave amplification via ECMI. Study of the nonlinear three-wave coupling process is rare.

Such a process satisfies the following matching conditions
\begin{equation}\label{eqcoups}
 \omega_1+\omega_2 = \omega_3, ~\vec k_1 + \vec k_2 = \vec k_3,
\end{equation}
where the subscripts ``1'' and ``2'' represent mother modes and the subscript ``3'' refers to the daughter mode. It is a necessary part of the PE theory \citep{Ginzburg&Zhelezniakov...1958}. See \cite{Thurgood&Tsiklauri...2015}, \cite{Henri.et.al.2019}, \cite{Ni.et.al.2020, Ni.et.al.2021}, and \cite{Li.et.al.2021} for the latest PIC simulation studies. A novel mechanism of harmonic maser emissions has been proposed which involves the coalescence of the Z mode \citep{Melrose.et.al.1984, Melrose...1991}. Yet, the coalescence conditions and wave saturation have only been investigated theoretically under simplifying assumptions.

In this study, we employ a fully kinetic electromagnetic PIC simulation in two spatial dimensions with three velocity components (2D3V). We aim to investigate the wave excitations and subsequent nonlinear interaction and saturation process, induced by energetic electrons with a loss-cone distribution in plasmas with a low value of $\wpe/\wce$. Such a simulation was not performed in former investigations to the best of our knowledge.

\section{Numerical model}

In the present study, the numerical simulation was performed with the Vector-PIC (VPIC) code developed and released by the Los Alamos National Laboratories (LANL). The code employs a second-order, explicit, leapfrog algorithm to update positions and velocities of charged particles, along with a full Maxwell description for evolution of electric and magnetic fields via a second-order finite-difference time-domain solver \citep{Bowers.et.al.2008b, Bowers.et.al.2008,Bowers.et.al.2009}. The simulation is performed using periodic boundary conditions, with a background magnetic field $\vec B_0 = B_0 \hat e_z$ and the wavevector ($\vec k$) in the $xOz$ plane.

To simulate the emission process under conditions of plasmas in solar active regions, we set $\wpe/\wce$ to be 0.25; the same values have been used by \cite{Yousefzadeh.et.al.2021}. The unit of length is taken to be the electron inertial length ($d_e = c/\wpe$) and the unit of time is taken to be the plasma response time ($\wpe^{-1}$). The simulation lasts for 8000~$\wpe^{-1}$, with a time step of $\Delta t = 0.024~\wpe^{-1}$. The domain of the simulation is $L_x = L_z = 1024 \Delta$, where $\Delta~(= 2.7~\lambda_\mathrm{DE})$ represents the grid spacing, and $\lambda_\mathrm{DE}$ is the Debye length of the background electrons. The wavenumber and frequency can be resolved in the ranges of $[-16, 16]$  $\wce/c$ and [0, 4.06] $\wce$, respectively; the resolution of the wavenumber is $\sim 0.031~\wce/c$, and the resolution in frequency is $\sim 0.003~\wce$ (with a time interval of 500~$\wpe^{-1}$). Charge neutrality is maintained initially; a realistic proton-to-electron mass ratio of 1836 is adopted. We included 2000 macroparticles for each species in each cell (nppc = 2000) with $\sim8.4\times10^9$ macroparticles in total. The simulation represents the largest one in terms of calculation resources ever conducted in studies on the same topic.

The background electrons are described by the following Maxwellian VDF
\begin{equation}\label{eq_maxw}
  f_0 = \frac{1}{(2\pi)^{3/2}v_0^3}\exp\left(-\frac{u^2}{2v_0^2}\right) ,
\end{equation}
where $u$ is the momentum per mass of particle $v_0 = 0.018c$ ($\sim$2 MK), and $c$ is the speed of light. Protons are described by the same VDF with the same temperature.

Energetic electrons are described by the following double-sided loss-cone VDF (see Figure~\ref{fig1}(a)),
\begin{equation}\label{eq_lossc}
   f_e(u,\mu)=A\ G(\mu)\exp\left(-\frac{u^2}{2v_T^2}\right),~  G(\mu)=1-\tanh\frac{|\mu|-\mu_0}{\delta} ,
\end{equation}
where $v_T = 0.2 c$ is the thermal velocity, $A$ represents the normalization coefficient, $\mu = u_\parallel/u$ is the cosine of the pitch angle, and $\delta~(= 0.1)$ determines the smoothness of the loss-cone boundary. The loss-cone angle is set to be ($\sim 30^\circ$) as $\mu_0 = 0.85$. The density ratio of energetic electrons to total electrons $n_e/n_0$ is set to be 10\%.

\section{Simulation results}

We present the evolution of the VDF, the wave-field energies, and the dispersion analysis in the following subsection to show the amplification process of the plasma waves and to identify the wave modes. Next we introduce the properties of the waves, especially the harmonic emissions as the major result of this simulation.

\subsection{Overview of wave amplification}

During the simulation, the electron VDF remains symmetric, as seen from Figure~\ref{fig1}(a)--(d) and the accompanying movie. After $\sim$1000--2000$~\wpe^{-1}$ electrons distributed along the loss-cone boundaries diffuse obviously; at $\sim4000~\wpe^{-1}$, diffusions appear in areas with small $|v_\parallel|$, as indicated at by purple arrows in panels (b)--(c). At the end of the simulation a major part of the original loss cones are filled up (panel (d)).

Figure~\ref{fig1}(e) presents the energy profiles of various field components, normalized to the initial kinetic energy of the total electrons ($E_\mathrm{k0}$). After 1000~$\wpe^{-1}$, the energies of most wave-field components (except $E_\mathrm{z}$) rise sharply. After 2500--4000~$\wpe^{-1}$, most components reach the corresponding saturation levels of energy.

At the end of the simulation ($t = 8000~\wpe^{-1}$), the energies of $E_x$ and $E_y$ reach $\sim 2\times 10^{-4}~E_\mathrm{k0}$, the energy of $B_y$ reaches $\sim 7\times10^{-5}~E_\mathrm{k0}$, and the energies of $E_z$, $B_y$, and $\Delta B_z$ reach $\sim 4\times10^{-5}~E_\mathrm{k0}$. The energy profiles of $E_x$ and $E_y$ are consistent with each other, following the relative decline of the kinetic energy of the total electrons ($-\Delta E_\mathrm{k}$). During the simulation, about $5\times 10^{-4}$ of $E_\mathrm{k0}$ are converted into wave energies.

We present the distributions of the maximal wave energies in the wavevector $\vec k\ (k_\parallel,\ k_\perp)$ space for electric fields over 7000--8000~$\wpe^{-1}$ in Figure~\ref{fig2}. As displayed in panels (a)--(c), three features with significant wave enhancement exist: (1) arc-shaped patterns with $-1<k_\perp <1~\wce/c$ (in $E_x$, and $E_y$), (2) vertical enhancement with  $|k_\parallel| \sim 0.3~\wce/c$ in $E_x$, and (3) growth in discrete regions with $|k_\perp|\sim 2~\wce/c$ in $E_y$. The enhanced waves can be identified as X1, Z mode, and second-harmonic emissions, respectively, according to the dispersion analyses (Figure~\ref{fig3}).
We note that waves at higher harmonics are not amplified significantly, thus not discussed further. We mainly analyze the waves in the first quadrant since wave patterns are symmetric.

We suggest that the second-harmonic emissions are mainly in X mode (X2). Figure~\ref{fig2}(e)--(f) present the maximal wave energies in a small range of $\vec k$, showing that the harmonic emissions are mainly carried by $E_y$ rather than $E_z$, along oblique to quasi-perpendicular directions, and the intensity of $E_z$ at 90$^\circ$ is too low to be observed. This is consistent with the dispersion relation of the magnetoionic X mode. We then divide the X2 mode into two components: X2$^\mathrm{A}$ at quasi-perpendicular directions ($80<\theta \le 90^\circ$) and X2$^\mathrm{B}$ at oblique directions ($\sim 65^\circ$--$75^\circ$, as marked in Figure~\ref{fig2}).

The temporal evolution of the wave energies in $\vec k$ space is displayed in the accompanying animation of Figure~\ref{fig2}. The strongest mode X1 is amplified at the earliest stage and maintains a high level of intensity thereafter. After 2000~$\wpe^{-1}$, the Z, X2$^\mathrm{A}$, and X2$^\mathrm{B}$ modes grow significantly.

\subsection{Characteristics of the amplified waves}\label{sec:modes}

Below we exhibit detailed features of the amplified modes (X1, Z, X2$^\mathrm{A}$, and X2$^\mathrm{B}$; see Figures~\ref{fig2}, Figure~\ref{fig3}, and the accompanying animations). The frequency of X1 is slightly above the X mode cut-off frequency ($\omega_\mathrm{X} = 1/2 (\sqrt{\wpe^2+4\wce^2}+\wce) \approx 1.06~\wce$) at $\theta = 0^\circ$, and rises to above 1.1~$\wce$ with $\theta$ increasing to 55$^\circ$. The bandwidth of X1 reaches $\sim0.1~\wce$ with $\theta>50^\circ$.
The quasi-electrostatic Z mode is amplified with a fixed $|k_\parallel| \sim 0.3~\wce/c$. With the value of $|k_\perp|$ increasing from 2 to 12~$\wce/c$, the intensity declines gradually (see Figure~\ref{fig2}(d)). With increasing values of $\theta$ and $k$, its frequency $\omega$ remains unchanged, close to the resonant frequency of Z mode ( $\sim 1.025~\wce$, see Figure~\ref{fig3}(b)).

The harmonic emissions X2 propagate at discrete angles. With $83^\circ<\theta<86^\circ$, X2$^\mathrm{A}$ is amplified at 2.09 and 2.13~$\wce$, separately. With $\theta$ varying in the range of $[88^\circ,\ 90^\circ]$, the frequency of X2$^\mathrm{A}$ is $2.05~\wce$. The X2$^\mathrm{B}$ is observed at $\sim66^\circ$ (2.14~$\wce$), 70$^\circ$ (2.09~$\wce$), and 73$^\circ$ (2.05~$\wce$), respectively. The intensities are lower than those of the X1 and Z modes, but still much higher than the corresponding noise level.

To evaluate the intensities of the wave modes, we perform fast Fourier transform (FFT) analysis on the wave-field data every 100~$\wpe^{-1}$, and integrate the energies within the respective ranges of $\vec k$ and $\omega$.
 The energy of the Z mode is obtained with $2<|k_\perp|<10~\wce/c$, neglecting that with $|k_\perp|>10~\wce/c$.
According to the temporal profiles of the mode energies (Figure~\ref{fig1}(f)), we split the simulation into Stage~\Rone\ (0--2000~$\wpe^{-1}$), \Rtwo\ (2000--4000~$\wpe^{-1}$), and \Rthree\ (4000--8000~$\wpe^{-1}$), representing the linear, nonlinear, and saturation stages. During Stage~\Rone, the energy of the X1 mode grows most impulsively with a fitted growth rate $\Gamma \sim 10^{-3}~\wce$, following the profile of $-\Delta E_\mathrm{k}$. Later, X1 saturates with an energy of $\sim3.0\times10^{-4}~E_\mathrm{k0}$ after 2500~$\wpe^{-1}$. The energy of the Z mode also rises exponentially in Stage~\Rone\ with a lower $\Gamma~(\sim5.2\times10^{-4}~\wce)$. In Stage \Rthree{}, the Z mode grows slowly, reaching a saturation level of $\sim4.7\times10^{-5}~E_\mathrm{k0}$ at the end.

X2$^\mathrm{A}$ grows at the beginning of Stage \Rtwo{}, with $\Gamma\sim4.5\times 10^{-4}~\wce$, comparable with that of the Z mode. The energy growth of X2$^\mathrm{B}$ gets significant after 3000~$\wpe^{-1}$ with $\Gamma \sim 2.2\times10^{-4}~\wce$. At the end of Stage \Rthree{}, X2$^\mathrm{A}$ and X2$^\mathrm{B}$ saturate at $\sim1.1\times10^{-5}$ and $\sim4.9\times10^{-6}~E_\mathrm{k0}$. Both the energies of X2$^\mathrm{A}$ and X2$^\mathrm{B}$ oscillate with time, different from those of the X1 and Z modes.

\section{Analysis on the mechanism of wave amplification}

Below we analyze the resonance conditions of ECMI and the matching conditions of three-wave interaction so as to clarify the radiation mechanism.

\subsection{On the ECMI process}

The resonance condition of ECMI can be written as
\begin{equation}
  \omega - k_\parallel v_\parallel - n \frac{\wce}{\gamma} = 0, \label{eqresonance}
\end{equation}
where $\gamma$ refers to the Lorentz factor and $n$ refers to the harmonic number. With appropriate values of $\omega$ and $k_\parallel$, the equation can be used to plot the resonance curves in the $(v_\parallel,~v_\perp)$ space.

\begin{figure*}[ht]
 \centering
 \includegraphics[width=13cm]{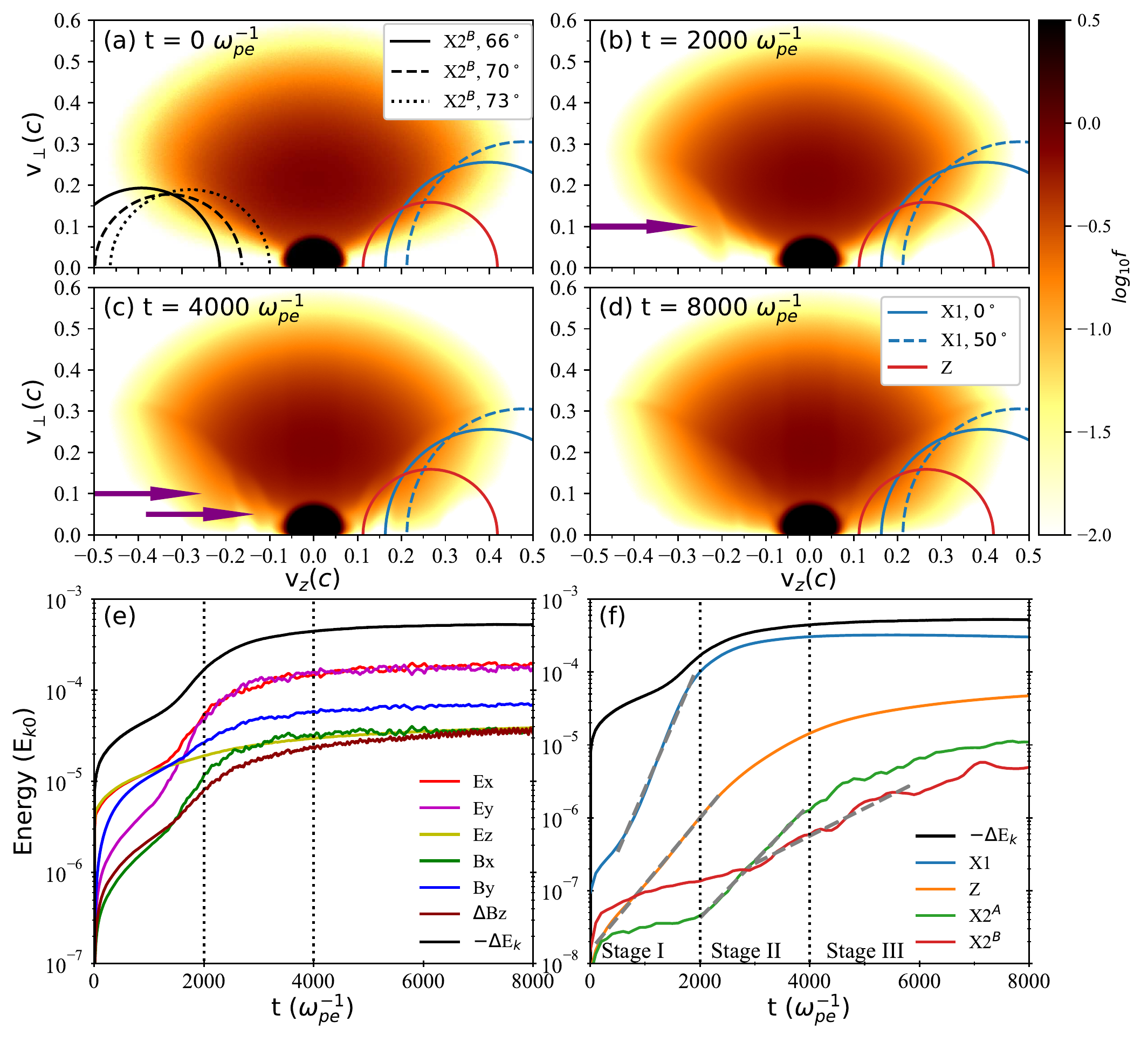}
  \caption{Panels (a)--(d): snapshots of the VDF of the simulation obtained at $t = 0 ~\wpe^{-1}$, $2000~ \wpe^{-1}$, $4000~ \wpe^{-1}$, and $8000~ \wpe^{-1}$. The purple arrows point to the areas of electron diffusion. The curves in blue, red, and black represent the resonance curves corresponding to the amplified waves in X1, Z, and X2$^\mathrm{B}$ modes. Parameters of the curves are listed in the last column of Table~\ref{table1}.
  Panels (e)--(f): temporal variations of energies of fluctuated fields ($E_x$, $E_y$, $E_z$, $B_x$, $B_y$, and $\Delta B_z$) and the amplified wave modes (X1, Z, X2$^\mathrm{A}$, and X2$^\mathrm{B}$). The energies are normalized to the initial kinetic energy of total electrons ($E_\mathrm{k0}$). The dark line refers to the relative decline of the kinetic energy of the total electrons ($-\Delta E_\mathrm{k}$). The gray dashed lines in panel (f) represent exponential fittings to the energy profiles, while the fitted growth rates are listed in Table~\ref{table1}. The dotted vertical lines denote the times of 2000 and 4000~$\wpe^{-1}$.
  ``X1'', ``Z'', ``X2$^\mathrm{A}$'', and ``X2$^\mathrm{B}$'' stand for the fundamental X mode, Z mode, second-harmonic X mode propagating along quasi-perpendicular directions, and that propagating along oblique directions, respectively.
  An animation of the panels (a)--(d) is available.}\label{fig1}
\end{figure*}

\begin{figure*}[ht]
 \centering
  \includegraphics[width=11cm]{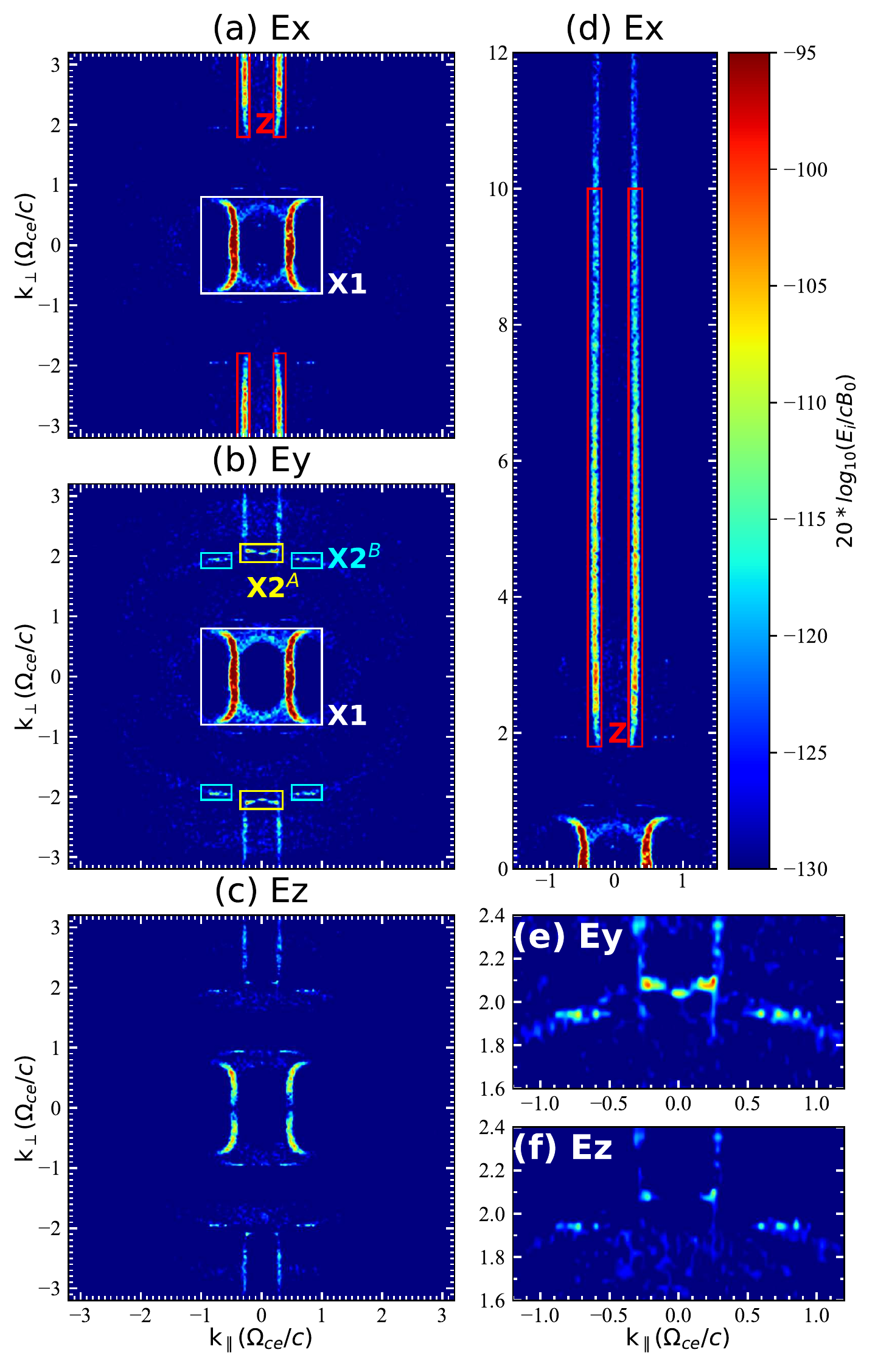}
  \caption{Maximal intensities of (a) $E_x$, (b) $E_y$, and (c) $E_z$ in the $\omega$ domain in the $k_\parallel$-$k_\perp$ space as shown by the color map of $20\log_{10}[(E_x,~E_y,~E_z)/(cB_0)]$, obtained over [7000, 8000]~$\wpe^{-1}$. Panels (d--f) present the same results within different ranges of $k_\parallel$ and $k_\perp$. All the images are normalized as shown by the color bar next to panel (d). The overplotted rectangles represent the areas used to obtain energies of the respective wave modes. See Figure~\ref{fig1}(f) for the obtained energy profiles. An animation of this figure is available. The movie begins with the interval of [0, 1000]~$\wpe^{-1}$, ending with the interval of [7000, 8000]~$\wpe^{-1}$. }\label{fig2}
\end{figure*}

\begin{figure*}[ht]
 \centering
  \includegraphics[width=13cm]{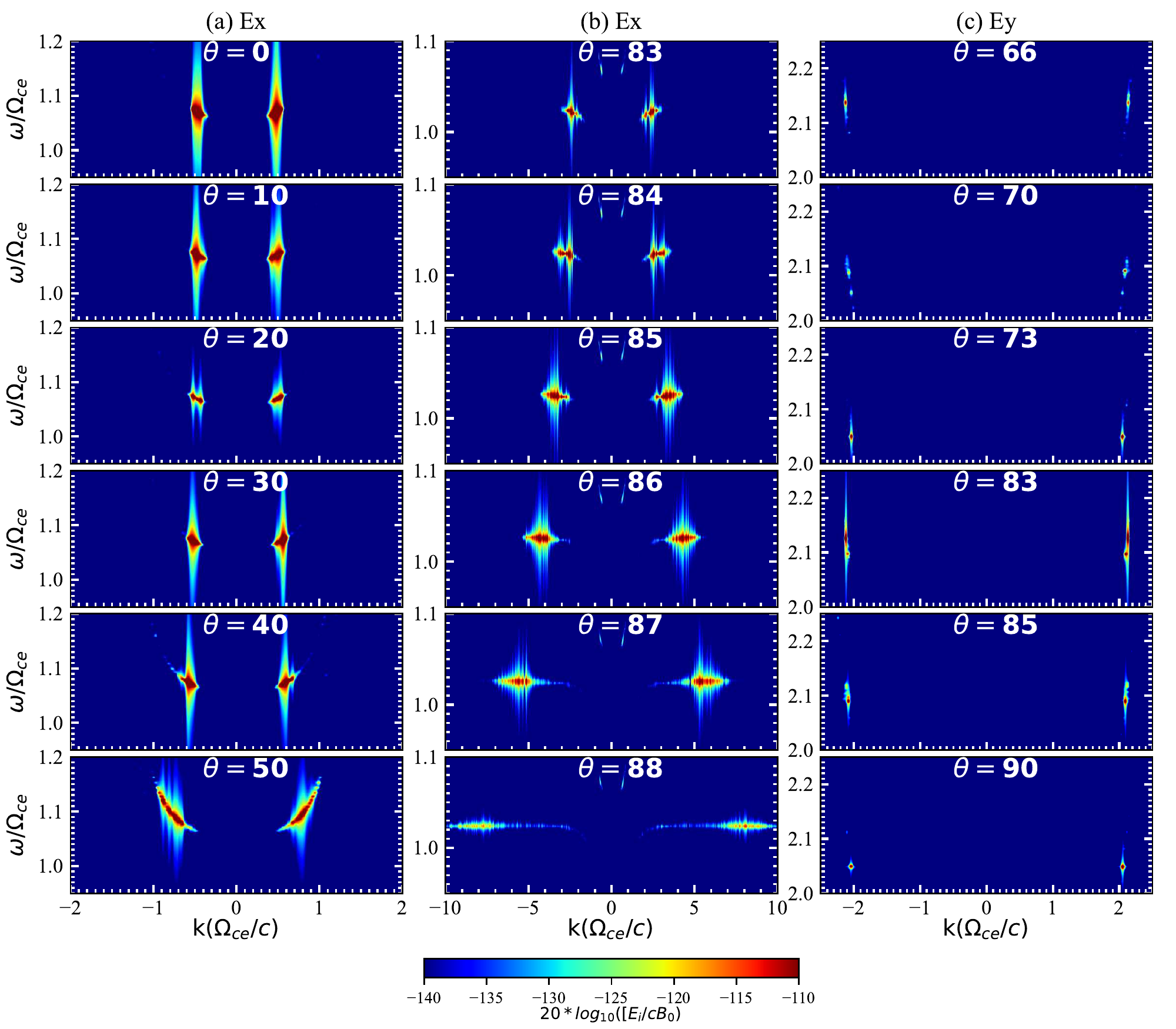}
  \caption{Wave dispersion diagrams of electric fields over 7500--8000~$\wpe^{-1}$ along different directions. Panels (a) and (b) display the diagrams of $E_x$ next to the frequency of $\wce$ along the parallel-oblique directions and quasi-perpendicular directions, respectively. Panel (c) displays the diagrams of $E_y$ with $2<\omega<2.25~\wce$. An animation of the figure is available. The movie begins with $\theta =0^\circ$, and advances 1$^\circ$ until ending at 90$^\circ$.}\label{fig3}
\end{figure*}

As illustrated above, values of both $\omega$ and $k_\parallel$ of the X1 mode vary slightly with $\theta$ (see Table~\ref{table1}). Thus the resonance curves of X1 at different directions are similar. See Figure~\ref{fig1} for two examples at $\theta = 0$ and 50$^\circ$.
 Note that for the amplified Z mode, values of $k_\parallel$ and $\omega$ are almost unchanged with $\theta$, yielding the same resonance conditions (see Figure~\ref{fig1} for the resonance curve). This explains why the Z mode is excited with an almost constant $k_\parallel$ (see Figure~\ref{fig2}(d)). All resonance curves of X1 and Z modes pass through the positive gradient regions of the VDF. At 2000 and 4000~$\wpe^{-1}$, electrons distributed along the resonance curves scatter significantly due to wave--particle interaction (see Figures~\ref{fig1}(b)--(c)).

For harmonic emissions propagating perpendicularly, the resonance condition can be simplified as
\begin{equation}\label{eqresperp}
  \omega = 2 \frac{\wce}{\gamma}
\end{equation}
which is satisfied only if $\omega<2~\wce$. However, the frequencies of X2$^\mathrm{A}$ are slightly higher than 2~$\wce$. By solving equation~\ref{eqresonance} with the obtained values of $\omega$ and $k_\parallel$, we found the resonance condition of X2$^\mathrm{A}$ cannot be satisfied. This means X2$^\mathrm{A}$ is not excited via ECMI.

We argue that the efficient excitation of X2$^\mathrm{B}$ should not be attributed to ECMI, though the resonance condition can be satisfied (see Figure~\ref{fig1}(a) for the resonance curves). Firstly, according to earlier studies, in plasmas with $\wpe/\wce<0.3$, the growth rate of harmonic emissions is generally low, which can hardly grow \citep{Winglee.et.al.1988,Aschwanden...1990}. Second, the ECMI tends to generate waves with continuous distributions in the $\omega$ and $\vec k$ domains, with resonance curves being clustered around adjacent locations of the phase space (such as X1 and Z modes). In the present work, the growth rate of X2$^\mathrm{B}$ is comparable with that of the Z mode and X2$^\mathrm{B}$ is excited at discrete frequencies within distinct areas in $\vec k$ space. Finally, as introduced in Section~\ref{sec:modes}, X2$^\mathrm{B}$ is amplified after 3000~$\wpe^{-1}$, while both the X1 and Z modes almost saturate with obvious diffusion of VDF. This also makes ECMI more difficult to amplify X2$^\mathrm{B}$.

In summary, the X1 and Z modes are amplified via ECMI, while X2$^\mathrm{A}$ and X2$^\mathrm{B}$ are not, according to the above analysis.

\subsection{On the wave--wave coalescence process}

\begin{figure*}[ht]
 \centering
  \includegraphics[width=12cm]{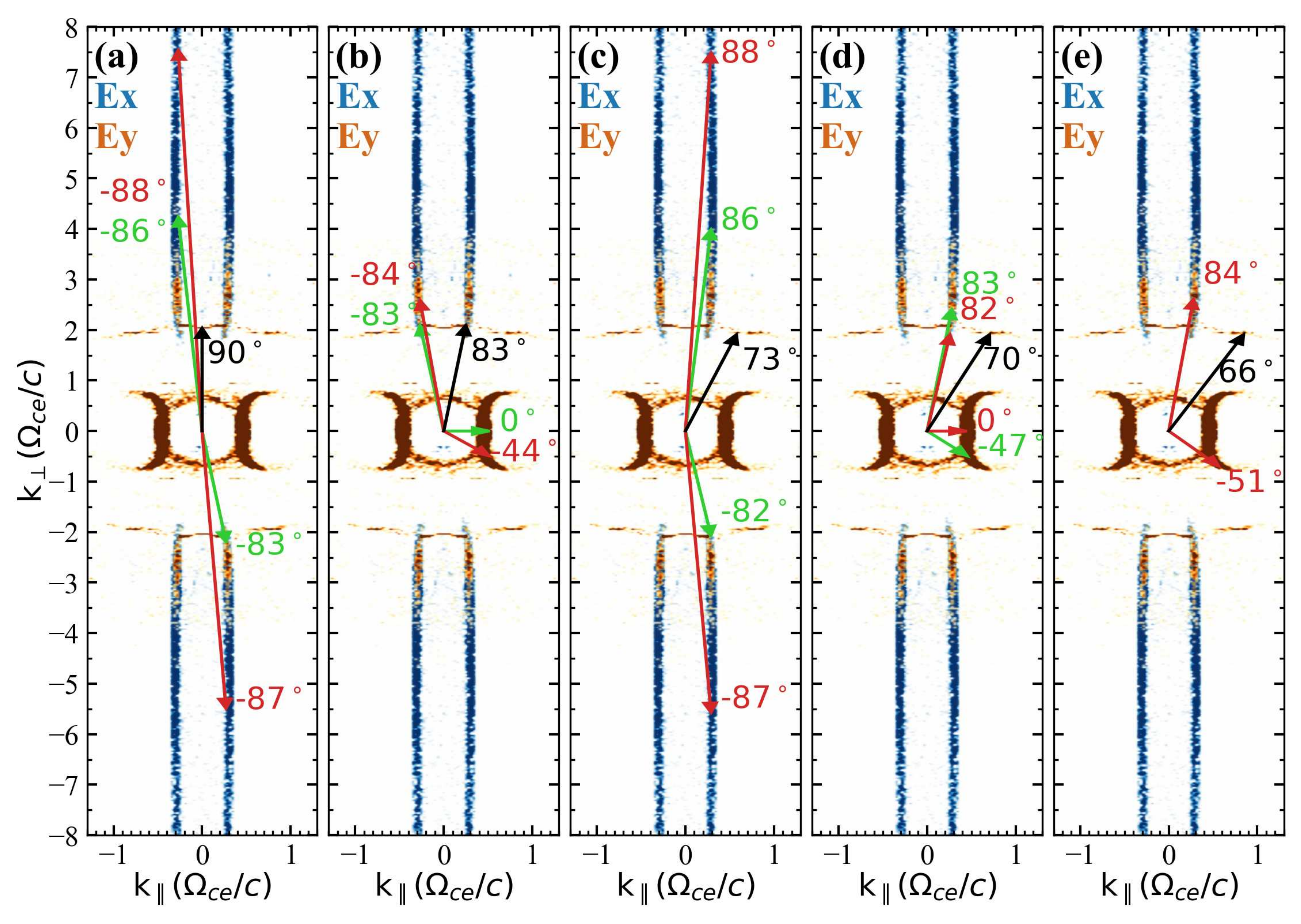}
  \caption{Maximal wave energies in $\vec k$ space over 7000--8000~$\wpe^{-1}$; blue and brown represent results in $E_x$ and $E_y$ components. Possible conditions of wavevectors matching the coalescence condition are overplotted as arrows in the same color. Black arrows represent the wavevector of the target mode to be generated. The values of $\theta$ of each wavevector is marked beside.}\label{fig4}
\end{figure*}

We then consider the wave--wave coalescence process to generate X2$^\mathrm{A}$ and X2$^\mathrm{B}$. It requires the matching conditions (equation~\ref{eqcoups}) to be met.
Possible candidates are plotted in Figure~\ref{fig4}. Panel (a) presents two sets of examples with almost counter-propagating Z modes ($k_\parallel \sim \pm 0.3~\wce/c$, $\omega \sim 1.025~\wce$, see Table~\ref{table1}) to generate X2$^\mathrm{A}$ at 90$^\circ$ ($k_\parallel \sim 0~\wce/c$, $\omega \sim 2.05~\wce$).
 As for X2$^\mathrm{A}$ at quasi-perpendicular directions $(83^\circ$--86$^\circ$), its wavevector ($[k_\parallel,~k_\perp]$ $\sim[0.25,~2.1]~\wce/c$) can be given by the vector sum of Z and X1 wavevectors ($k_\parallel \sim -0.3$ and $0.55~\wce/c$; see arrows in panel (b) for two sets of examples). The frequency of X1 ranges from 1.06 to 1.14~$\wce$, thus its coalescence with Z mode ($\sim1.025~\wce$) could account for the amplified X2$^\mathrm{A}$ (2.09 and 2.13~$\wce$). With the values of $k_\parallel$ and $\omega$ changing minorly, the mother Z and X1 modes within a wide range of $k_\perp$ can satisfy the matching condition as long as the difference in $k_\perp$ matches $k_\perp$ of the daughter wave.

Similarly, the coalescing processes of $\mathrm{Z}+\mathrm{Z}\rightarrow \mathrm{X2}^\mathrm{B}$ and $\mathrm{Z}+\mathrm{X1}\rightarrow \mathrm{X2}^\mathrm{B}$ can occur for mother modes (Z and/or X1) in the first and the fourth quadrant. The parameters for the matching conditions (Equation~\ref{eqcoups}) are listed as follows: (1) X2$^\mathrm{B}$ at 73$^\circ$ ($k_\parallel \sim 0.6~\wce/c$, $\omega\sim2.05~\wce$) can be generated by coalescing of Z modes with $k_\parallel \sim 0.3~\wce/c$ and $\omega\sim 1.025~\wce$ (see arrows in panel (c)); (2) X2$^\mathrm{B}$ at 70$^\circ$ ($k_\parallel\sim 0.7~\wce/c$, $\omega\sim2.09~\wce$) by coalescing of Z ($k_\parallel \sim 0.3~\wce/c$, $\omega\sim1.025~\wce$) and X1 ($k_\parallel \sim 0.4~\wce/c$, $\omega\sim 1.07~\wce$) modes (panels (d)); and (3) X2$^\mathrm{B}$ at 66$^\circ$ ($k_\parallel \sim 0.87~\wce/c$, $\omega\sim2.14~\wce$) by coalescing of the same Z mode with X1 at directions of $|\theta|>48^\circ$, requiring large values of $k_\parallel$ ($\sim0.57~\wce/c$) and $\omega$ ($\sim1.12~\wce$) of X1 (panel (e)).

In addition, the following signatures also support the proposed coalescing processes. (1) X2$^\mathrm{A}$ and X2$^\mathrm{B}$ are amplified at separate frequencies and within discrete areas of $\vec k$ space, in line with the limited varying ranges of $\omega$ and $k_\parallel$ of the Z and X1 modes. (2) X2$^\mathrm{A}$ and X2$^\mathrm{B}$ grow later than X1 and Z, and the X2$^\mathrm{A}$ energy profile is similar in increasing trend with that of the Z mode. We found that the energy ratio of X2, including X2$^\mathrm{A}$ and X2$^\mathrm{B}$), to the Z mode is $\sim30\%$, and to X1 $\sim 5\%$, indicating efficient energy conversion of the three-wave resonant coupling process.

\section{Conclusions and Discussion}

In the present study, we investigated the harmonic emissions generated by electrons with a loss-cone distribution in plasmas with $\wpe/\wce$ as low as 0.25, using a fully kinetic electromagnetic PIC simulation. As known X1- and Z-mode waves are amplified during the early stage of the simulation via the ECMI, it is new in this study that the second-harmonic X-mode emissions (X2) can be generated later in restricted angles along the perpendicular and oblique directions. It appears that the X2 emissions are mainly caused by the resonant wave coupling process of almost counter-propagating Z modes ($\mathrm{Z}+\mathrm{Z}\rightarrow \mathrm{X2}$) and the $\mathrm{Z}+\mathrm{X1}\rightarrow \mathrm{X2}$ process. The total energy of X2 emissions can reach $\sim1.5\times 10^{-5}~E_\mathrm{k0}$, 30\% of that of Z mode. These results shed light on the long-standing problem of how harmonic emissions can be amplified efficiently by the loss-cone distribution in plasmas with a low value of $\wpe/\wce$.

This new finding owes to the simulation performed within a large-scale two-dimensional domain with a long enough duration, which allows the occurrence of the nonlinear wave--wave coalescence. In addition, with a huge amount of macroparticles the noise level is lowered and the harmonic emissions are well resolved.

Since harmonic EMCE at $\omega > 2~\wce$ is more likely to escape from the solar corona without passing through the second-harmonic layer, the simulated process can resolve the escaping difficulty of the ECME theory for solar radio bursts. In the simulation, more than $10^{-5}$ of $E_\mathrm{k0}$ can go to escaping emissions, with narrow bandwidth and propagation angles. This accounts for emissions with high brightness temperatures and explains why only 2\% of the Hard X-ray bursts correlate with radio spikes \citep[due to limited propagation angles; see, e.g.][]{Guedel.et.al.1991}.

The energy of the X2 emissions reaches $\sim30\%$ of the energy of the Z mode, indicating an efficient process of energy conversion via wave--wave coalescence. This is likely because that the coalescence could occur for X1 and Z modes in a wide range of $k_\perp$. To verify the occurrence of such a process and examine the efficiency of wave coupling, further studies should be carried out, as done by \cite{Ni.et.al.2021} with the wave-pumping method of PIC simulations. Simulations with varying $\wpe/\wce$ and with different VDFs should also be performed to explore more properties of ECMI and wave coalescence.

\begin{acknowledgments}
This study is supported by NNSFC grants (11790303 (11790300), 11973031, and 11873036). The authors acknowledge the Beijing Super Cloud Computing Center (BSCC, URL: http://www.blsc.cn/) for providing HPC resources, and the open-source VPIC code provided by LANL. We thank Dr. Jeongwoo Lee (NJIT) for helpful suggestions.
\end{acknowledgments}


\begin{table}[ht]
\centering
       \begin{tabular*}{\hsize}{p{0.05\textwidth}  p{0.07\textwidth}  p{0.10\textwidth}  p{0.10\textwidth}  p{0.10\textwidth}  p{0.12\textwidth}p{0.05\textwidth} p{0.05\textwidth} p{0.05\textwidth} p{0.05\textwidth}} 
         \hline\hline
         \multirow{2}{*}{Mode} &  \multirow{2}{*}{$\theta$} &  \multirow{2}{*}{$\omega (\wce)$} &  \multirow{2}{*}{$k (\wce/c)$} &  \multirow{2}{*}{$\Gamma$ ($\wce$)}&  \multirow{2}{*}{Energy ($E_\mathrm{k0}$)} &\multicolumn{4}{c}{Resonance curves} \\

                                                       & & & & & & $\theta$ & n & $\omega (\wce)$ & $k_\parallel (\wce/c)$ \\
         \hline
         \multirow{2}{*}{X1} & \multirow{2}{*}{0--55$^\circ$} & \multirow{2}{*}{1.06--1.14} & \multirow{2}{*}{0.4--0.9} & \multirow{2}{*}{1.0E-3} & \multirow{2}{*}{3.0E-4} & 0$^\circ$ & 1 & 1.06 & 0.45\\
                                                 & & & & & &  50$^\circ$ & 1 & 1.1 & 0.58 \\
         Z & 82--89$^\circ$ & 1.025 & 2--10 & 5.2E-4  & 4.7E-5 &86$^\circ$ & 1 & 1.025 & 0.28 \\
         \multirow{3}{*}{X2$^\mathrm{B}$}& 66$^\circ$ & 2.14 & 2.14 & \multirow{3}{*}{2.2E-4} &\multirow{3}{*}{4.9E-6}  & 66$^\circ$ & 2 & 2.14 & -0.87\\
                                                                    & 70$^\circ$ & 2.09 & 2.09 & & & 70$^\circ$ & 2 & 2.09 & -0.71 \\
                                                                    & 73$^\circ$ & 2.05 & 2.05 & & & 73$^\circ$ & 2 & 2.05 & -0.60 \\
         \multirow{2}{*}{X2$^\mathrm{A}$}& 82--85$^\circ$ & 2.09, 2.14 & 2.09, 2.14 & \multirow{2}{*}{4.5E-4} &\multirow{2}{*}{1.1E-5}&  & & & \\
                                                & 88--90$^\circ$ & 2.05 & 2.05 &  &   & & &\\

         \hline

       \end{tabular*}
       \caption{Parameters of the amplified modes. The last column present the corresponding parameters of the resonant curves plotted in Figure~\ref{fig1}.}\label{table1}
\end{table}

\end{CJK*}
\end{document}